%% file: main.tex
\documentclass[sigconf]{acmart}

\usepackage{booktabs} 
\usepackage{alg}
\usepackage{balance}

\acmConference[WWW 2018]{The 2018 Web Conference}{April 23--27, 2018}{Lyon, France}
\acmBooktitle{WWW 2018: The 2018 Web Conference, April 23--27, 2018, Lyon, France}

\begin{document}

\title{Crowd-based Multi-Predicate Screening of Papers in Literature Reviews}

\author{Evgeny Krivosheev}
\affiliation{%
  \institution{University of Trento, Italy}
}
\email{evgeny.krivosheev@unitn.it}

\author{Fabio Casati}
\affiliation{%
  \institution{University of Trento, Italy}
  \institution{Tomsk Polytechnic University, Russia}
}
\email{fabio.casati@unitn.it}

\author{Boualem Benatallah}
\affiliation{%
  \institution{University of New South Wales, Australia}
 }
\email{boualem@cse.unsw.edu.au}

%
%




\input{abstract}
\maketitle

\input{introduction}

\input{related}

\input{models}

\input{algorithms}

\input{experiments}

\input{conclusion}

\bibliographystyle{ACM-Reference-Format}
\balance
\bibliography{cs.bib}

\end{document}

%% file: abstract.tex
\begin{abstract}
Systematic literature reviews (SLRs) are one of the most common and useful form of scientific research and publication. 
Tens of thousands of SLRs are published each year, and this rate is growing across all fields of science. 
Performing an accurate, complete and unbiased SLR is however a difficult and expensive endeavor. 
This is true in general for all phases of a literature review, and in particular for the  paper screening phase, where authors filter a set of potentially in-scope papers based on a number of \textit{exclusion criteria}. 
To address the problem, in recent years the research community  has began to explore the use of the crowd to allow for a faster, accurate, cheaper and unbiased screening of papers. Initial results show that crowdsourcing can be effective, even for relatively complex reviews.

In this paper we derive and analyze a set of strategies for crowd-based screening, and show that an adaptive strategy, that continuously re-assesses the statistical properties of the problem to minimize the number of votes needed to take decisions for each paper, significantly outperforms a number of non-adaptive approaches in terms of cost and accuracy. 
We validate both applicability and results of the approach through a set of crowdsourcing experiments, and discuss properties of the problem and algorithms that we believe to be generally of interest for classification problems where items are classified via a series of successive tests (as it often happens in medicine). 

\end{abstract}

%% file: introduction.tex
\section{Introduction}
Systematic literature reviews (SLR) \cite{Grant2009ReviewTypes,khan2003five,henderson2010write} are reviews that follow a predefined process aimed at achieving transparency and impartiality with respect to the sources analyzed, minimizing distortions, biases, and conflicts of interest \cite{steinberg2011clinical}.
They are one of the most important form of publications in science \cite{Sun16-hcomp}, and are the basis for evidence-based practices and even government policies, from education to healthcare, as they pool results independently obtained from a number of research groups \cite{Haidich2010}. 
Recognizing their importance, the number of systematic reviews is growing steadily, with tens of thousands of publications per year in all fields. 

The cornerstone of transparency and impartiality in SLRs lies in a formalized paper selection process. 
This is typically formed by a stated scope and goal of the review (e.g., "study the effect of regular physical exercises on progress of dementia in older adults, focusing only on papers describing randomized controlled trials"), translated by the authors into a corresponding query (a boolean expression that includes relevant keywords) that retrieves candidate papers from a database such as Scopus.
To avoid missing papers, the query tends to be inclusive, which means that it returns hundreds or thousands of results \cite{Nguyen2015} that are later screened by researches based on predefined \textit{exclusion criteria} (e.g., "filter out papers that do not measure cognitive decline"), typically down to a few dozens.


While extremely useful, SLRs are very time consuming in terms of both effort and elapsed time, and this is true also for the paper screening phase \citep{coch_handbook2011,Sampson2008,Krivosheev_hcomp17}.
Furthermore, with hundreds of thousands of papers written every year, SLRs rapidly become outdated~\citep{wasted2016}, and although they should be updated periodically, the effort for doing so often represents a barrier \cite{Takwoingi2013}, so that it is not uncommon for reviews to miss 30\% or 40\% of relevant papers~\citep{wasted2016}. 

In this paper we explore the use of crowdsourcing in systematic reviews, and specifically in the filtering phase, where we screen candidate papers resulting from the initial literature search to identify papers to be included in the analysis.
This is an instance of finite pool classification  \cite{Nguyen2015} and crowd screening problems \citep{crowdscreen12} where we need to classify a finite set of objects while minimizing cost. 
The potential benefits of crowdsourcing here are in terms of a faster and cheaper screening (compared to screening by professionals) as well as increased transparency (process and votes can be made public if desired) and reduced risk of author bias. The crowd also brings diversity \cite{Weiss2016} and, as we experienced first hand, disagreement in the crowd may signal errors or ambiguities in the definition of exclusion criteria.
Research in this area is still in its infancy, although a set of recent initial efforts~\cite{Wallace2017crowdML,Krivosheev_hcomp17,Sun16-hcomp,Mortensen2016crowd} present very encouraging results in terms of both quality and cost reduction with respect to expert screening costs, and show feasibility of crowd-based screening in various domains, including healthcare.  


In the following we present a probabilistic model suitable for the  criteria-based screening of papers typical of SLRs and propose a set of strategies for crowd-based screening. 
Our main contribution consists in an adaptive crowdsourcing algorithm that significantly outperforms baselines.
The algorithm polls the crowd in small batches and estimates, at each iteration and \textit{for each item}, i) the criterion for which getting one more crowd vote on the paper can more efficiently lead us to a classification decision, and ii) whether we should give up trying to classify this item, recognizing that the crowd cannot efficiently reach a decision and therefore it should be left to the authors for expert screening.
This also means that the algorithm is robust to papers and criteria that are overly difficult for the crowd to classify, in that it does not needlessly spend money on them.  


The model is the result of many iterations and variations of experiments on commercial crowdsourcing platforms (Amazon Mechanical Turk (AMT) and \textit{CrowdFlower}\footnote{www.mturk.com and www.crowdflower.com}). We then performed additional experiments to validate the effectiveness of the strategies. 
While we present the results in the context of SLRs because we validated the model and findings for this case, we believe that results can be generally of interest for classification problems where items are classified via series of successive tests, as it often happens in medicine, as well as for finite pool classification problems and crowd-based query optimization, where the crowd evaluates predicates (analogously to our exclusion criteria) that filter a set of tuples to compute the query results.

%% file: related.tex
\section{Related work}
Our work builds on approaches in crowdsourcing in SLR but also more generally on works on crowd-based classification. 

\textbf{Crowdsourcing in Systematic Reviews.}
Recently, Brown and Allison~\cite{brown_crsliterature2016} used crowdsourcing to, among other tasks, classify 689 abstracts based on a set of criteria using AMT.
Authors report agreement in 75\% of the abstract, based on two raters, and a third rater is used to break the tie in case of disagreement. The paper does not discuss optimal crowdsourcing strategies or algorithms to minimize errors, but points to the potential of crowdsourcing in analyzing literature.  

Mortensen and colleagues crowdsourced paper screening~\cite{Mortensen2016crowd} in four literature reviews, each with several criteria.
Their aim was to explore feasibility and costs of crowdsourcing and they address the problem by measuring workers agreement in a set of tasks run on AMT for papers in the medical field.
Their work differs from ours in that it does not propose algorithms to identify optimal crowdsourcing strategies. 
However, it contains interesting observations related to the importance of task design, to the cost-effectiveness of crowdsourcing even when the task is not optimized, and to the high degree of variability in workers's agreement from paper to paper and criteria to criteria (Fleiss' Kappa ranging from 0.5 to -0.03). This is consistent with our own studies (our papers are in a different scientific area) and we exploit this variability to optimize the cost/error tradeoff. 

Krivosheev and colleagues~\cite{Krivosheev_hcomp17} also present a model and strategies for crowdsourcing SLR. An interesting aspect of the model and approach here is that the authors model cost and loss (error) resulting from crowdsourcing task, attempt to estimate them at the start, and provide authors with a price/error trade-off that can be used to decide how much to invest in the task.  
We borrow several concepts from this work, such as the ability to provide an estimate and a set of alternatives to SLR author, although the model of this paper is limited to screening based on one criterion.

Nguyen et al. \cite{Nguyen2015} adopt a mixed crowd+expert+machine learning approach with an active learning classifier, where papers to be labeled are iteratively chosen to minimize overall cost and loss, by comparing estimated loss of crowd classification versus expert classification. This paper is part of a trend trying to leverage AI in literature reviews, which we do not discuss further as it is not the focus of this paper. 



In general all papers reports positive results and complement them with insights and guidelines for task design of even for the design of a dedicated crowdsourcing platform for SLR \cite{Weiss2016,brown_crsliterature2016} as well as investigate the use of crowd for other phases of interest for SLR such as information extraction \cite{Sun16-hcomp}. Interestingly, the only exception is represented by a study performed with medical students as screeners rather than online crowd workers, which reports rather poor accuracy \cite{Ng2014}. 

From these studies we also learn that workers' accuracy vary across criteria, which points to the need of adapting to the characteristics of each SLR, criterion, and crowd.
Indeed, one of the main differences of our approach lies in the ability to focus the crowd on "low hanging fruits", that is, items and criteria that are statistically more efficient from the perspective of correctly excluding papers.
 
Although not focused on paper screening, we also mention a fascinating analysis by Law and colleagues trying to understand under which conditions do researchers resort to crowdsourcing \cite{Law_2017_uncertainty}. 
Among the many interesting considerations lies the observation that crowdsourcing is viable only if both authors \textit{and reviewers} find it acceptable.
Paper screening in SLRs seem to fit the requirements for being acceptable by authors but it is equally important for the scientific community to provide solid evidence of the quality of crowdsourced screening if we want it to be accepted by reviewers - especially in fields where SLRs may form the base of policies and practices.

\textbf{Crowdsourced Classification}
The problem discussed here is an instance of a finite pool classification problem \cite{Nguyen2015} and specifically of crowdsourcing-based classification.
This problem has been studied for hundreds of years now, dating back at least to the end of the 18$^{th}$ century, when the  Marquis de Condorcet presented his \textit{Jury Theorem}\footnote{http://www.stat.berkeley.edu/~mossel/teach/ SocialChoiceNetworks10/ScribeAug31.pdf}, stating that if each juror in a jury has an error rate lower than 0.5 and if \textit{guilty} vs \textit{innocent} votes are independent, larger juries reach more accurate results, and approach perfection as the jury grows.

From there, researchers from the AI, database, and human computation communities have proposed many of classification algorithms, mostly based on variations of majority voting where votes are counted differently based on estimated worker's accuracy. 
The seminal work of Dawid and Skene \cite{DawidSkene_Confusion} and refinement by, among others, Whitehill \cite{whitehill2009whose}, Dong et al \cite{Dong2013}, Li et al \cite{Li_error_13}, and Liu et al \cite{LiuWang_truelabel,liu2013scoring} model workers' accuracy - often with a confusion matrix - and then adopt variants of Expectation Maximization \cite{EM_1977} to iteratively refine prior estimates of workers' accuracy and of labels. 
Approaches based on spectral methods \cite{karger2011iterative} and maximum entropy \cite{Zhou_spectral_13} have also been proposed, and belief propagation has  been  recently shown  \cite{Ok_belief_13} to be optimal under certain assumptions. 

Prior work also addresses the issue of optimizations in terms of costs for obtaining labels and techniques to reduce cheating \cite{smyth1995inferring,karger2011budget,hirth2013analyzing,eickhoff2013increasing,hirth2011cost}. For example, Hirth and colleagues \cite{hirth2013analyzing} recommend specific cheating detection and task validation algorithms based on the cost structure of the task. 

We build over many of these approaches, and in fact we adopt prior art algorithms for estimating workers' accuracy and for assigning labels. 
Although classification algorithms are central to our overall problem, to a large extent they are for us a swappable component: Our goal is to, given a task design and a classification algorithm, identify how to efficiently query the crowd to minimize the number of labels needed to achieve the desired precision and recall in screening problems.

%% file: models.tex
\section{Model and Objective}
We model the SLR screening problem as a set of papers (items) $I$ to be classified by the screening phase as \textit{included} (in scope) or excluded based on a set of  exclusion criteria (predicates) $C=\{c_1,c_2,...c_m\}$.
A paper is excluded if at least one exclusion criterion applies, otherwise it is included. A typical SLR screens hundreds or thousands of papers with a handful of exclusion criteria. We focus on screening based on title and abstract, which is a classical first step screening, consistent with SLR guidelines~\cite{moher2009preferred}.

In a crowdsourcing approach, we ask each crowd worker to look at one or more pairs $(i,c)$ and state if  exclusion criteria $c$ applies to paper $i$. 
Following the mentioned literature, we model a worker's accuracy with a confusion matrix $A_{c,w}$ defining the probability of making correct and wrong classifications for each criterion $c$, thereby allowing us to model different accuracies when the true label is inclusion vs exclusion.  
Criteria can differ in \textit{difficulty}. Some are easier to assess than others. Following Whitehill \cite{whitehill2009whose}, we model difficulty as a positive real number $d_c$ that, given an expected accuracy $\alpha_w$ of a worker $w$, skews the accuracy as follows: $\alpha_{c,w} = 0.5+ (\alpha_w-0.5)*e^{-d_c}$. As the difficulty $d_c$ grows, $\alpha_{c,w}$ goes to 0.5, corresponding to random selection, which we consider to be the lowest accuracy level\footnote{In this paper we do not consider the problem of accuracies below random, but we stress that they can occur in rare cases, for example if criteria are erroneously specified.}. 
Each criterion also has a \textit{power} (also called selectivity) $\theta_c$, defined as the percentage of papers to which the criterion applies (and hence need to be excluded). 
For each SLR and criterion, both accuracy and power are unknown a priori. 

We assume the adoption of a general purpose crowdsourcing system with limited control on the kind of crowd we attract but with a near infinite pool of workers. 
We can however test workers by providing a number $N_t$ of test questions (with gold answers provided by SLR authors), and count as valid only votes of workers who pass the test, thereby exercising some control over worker's accuracy (at a cost, as we specify later).

A crowdsourcing \textit{strategy} is a set $K$ of runs, where each run $R^k$ collects  $J^k_{i,c}$ votes for  criterion $c$ on item $i$. A run may seek votes on all criteria and all papers, or focus on a subset (that is, $J^k_{i,c}$ might be 0 for some items). 

Tasks also have a cost, which is the unit cost $UC$ for a (non-test) vote multiplied by the number of votes obtained. Although many systems allow not to pay for test answers, consistently with \cite{Krivosheev_hcomp17}, we believe it is fair and ethical to also pay for test questions for workers who pass them. Furthermore, placing unreasonably many test questions is likely to result in low reputation scores for us and hence in our ability to crowdsource. 
Concretely, this translates into considering a price per label $PPL$ as follows ($N_{l}$ is the number of valid judgments that a worker gives on non-test papers)

\begin{equation}\label{ppp}
PPL = UC \cdot \frac{N_{l}+N_t}{N_{l}}
\end{equation}

The correction factor approaches 1 when $N_l$ is large compared to $N_t$. In practice our control on $N_l$ can be limited by many factors (also depending on the crowdsourcing platform policies), such as dropouts, the presence of many concurrent workers that exhaust the available tasks, and more. 
We observe that tests are for us "simply" a knob to turn when trading costs for accuracy. Any other knob that accomplishes the same effect can be equivalently used in what follows. 

In terms of outcome, key measures are the precision and recall of papers to exclude. 
We also borrow the concept of loss function from \cite{Krivosheev_hcomp17,Nguyen2015} because it summarizes well the subjective perspective of the SLR authors. The  $loss = lr*FE + FI$ is represented by the sum of false inclusions $FI$ (papers that survived the screening phase but that should have been excluded instead) and false exclusions $FE$ (filtered-out papers that should have instead been left in), where $FE$ are weighed by a loss ratio $lr$ denoting that false exclusion are $lr$ times more "harmful" than false inclusion (filtering out a paper is often considered a much more serious error than a false inclusion which "simply" requires extra work by the authors).
The loss ratio is the only parameter we ask the authors to set.

Many variations of the model and of loss function are possible, but these suffice for our purposes. 
Given the model, our objective is to identify and evaluate a set of efficient crowdsourcing strategies for each SLR that correspond to estimated pareto-optimal price/loss curves. With infinite money we can always arrive at a perfect classification (if workers' accuracy is above random and votes are independent), but the challenge is to classify efficiently and at a price/loss point that is acceptable to authors, who decide what price they are willing to pay and which loss they can tolerate. Based on this preference, the algorithm should set the relevant parameters of the crowdsourcing tasks and classification function. We next discuss how this can be done.



%% file: algorithms.tex
\section{Algorithms}
\subsection{Baseline single-run algorithms}
Our set of baseline algorithms follows the methods applied in recent literature for crowdsourced classification in finite pool contexts and SLRs in particular. 
Specifically, as we are in the presence of incomplete information (we know neither the classification of the papers nor the accuracy of the workers), we leverage approaches such as TruthFinder \cite{Dong2013} and Expectation Maximization (EM,  ~\cite{DawidSkene_Confusion}) to iteratively refine estimates of accuracy and class until convergence. 
In addition, simple majority voting is also commonly used as its performances are actually reasonable in finite pool classification~\cite{Nguyen2015}.

Applying them to our problem, we proceed in a single run where we ask each worker to vote on all criteria $C$ for a set of papers. Each worker provides at most $N_l$ labels, and we collect $J$ votes per criteria and per paper. Classification proceeds by evaluating each criterion $c \in C$ on each paper $i$ and, based on the responses received, estimating with one of the mentioned algorithms the probability $P (i \in OUT_c)$ that paper $i$ is classified as out by criterion $c$.

Once we have probabilities for each criterion,  we compute the probability $P(i \in OUT)$ that a paper $i$ should be excluded as the probability that at least one criterion applies (we assume criteria application is independent): 
\begin{equation}
	\begin{aligned}
		P(i \in OUT) = 1 - \prod_{c \in C} P(i \in IN_c)
	\end{aligned}
	\label{formula:p-out}
\end{equation}

The loss ratio skews our classification decision to err on the side of inclusion (for $lr$>1).
The expected loss per paper we suffer for an erroneous inclusion of a paper $i$ is $P(i \in OUT)$, while for an erroneous exclusion it is $lr \cdot (1-P(i \in OUT))$.
This means that our threshold for classifying a paper as OUT is when these quantities are the same, that is, $P(i \in OUT) = lr/(lr+1)$. 

\begin{figure}[htb]
	\centering
		\includegraphics[width=0.5\textwidth]{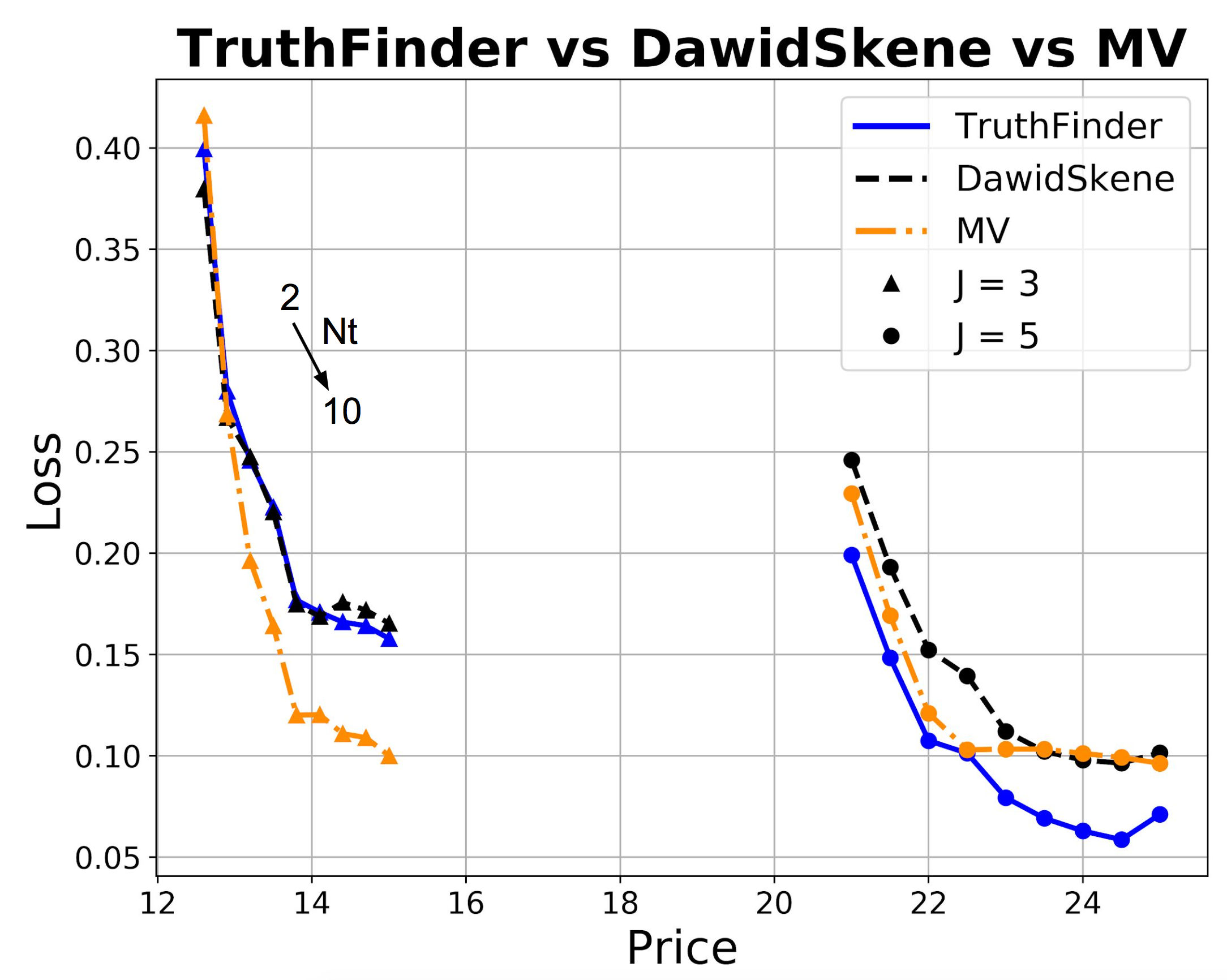}
		\caption{Performance of classification algorithms. Simulation with 1000 papers, four criteria of power $=[c1=0.14, c2=0.14, c3=0.28, c4=0.42]$, $Nt=[2,3,..,10]$, $lr=5$. Workers are assumed to be cheaters with probability 0.3, and the rest has uniform accuracy in (0.5-1). Accuracy on OUT papers are 10\% higher, as seen in experiments.} 
\label{fb}
\end{figure}

Altering the number of votes per worker $N_l$, votes per item $J$, and number of tests $N_t$ will modify the expected price and loss.
More tests ideally lead to more accurate workers, more labels mean more accurate classification, and  more votes per person enable a more accurate estimation of a worker's accuracy. 
To analyze price vs loss, we simulate the behavior of the model with various values of $N_l, J, N_t$, and apply EM, TruthFider (TF) or Majority Voting (MV) to classify papers, and compute the estimated loss. 
Since values of $N_t$ and $J$  correspond to a cost, we can also get the price tag corresponding to this loss.
Out of this set of price/loss points, we can take the pareto-optimal ones and plot them so that authors can decide which one best fits their needs.
As discussed there are cost penalties and practical constraints that do not allow us to set these parameters to arbitrarily high values, and values of $N_t$ and $J$ above 10 do not generate significant improvements \cite{Krivosheev_hcomp17}, so the number of reasonable alternatives is fairly small. 
To simulate the data we need either to make assumptions on the crowd accuracy as well as on criteria power and difficulty, possibly based on prior knowledge, or to estimate these parameters \cite{Krivosheev_hcomp17} by crowdsourcing labels for a few papers (fifty papers already enable a good estimate as shown later).
Figure~\ref{fb} shows the results of applying the three mentioned algorithms for 3 and 5 labels per item and criterion (the caption describes simulation parameters). The impact of choosing a specific algorithm is relatively small with the exception of MV performing better when labels per paper and per worker are few, which is a known behavior \citep{labelquality_hcomp17}.
The dots represent different number of tests (from 1 to 10) and the arrows shows the direction of growth, from top-left to bottom-right.  
Some points are Pareto-optimal, so in an interaction with SLR authors we would only show those points and ask for the preferred loss/price point.

The results vary slightly if the parameters of the problem and algorithms are different (such as different power,  difficulty distributions across criteria, proportion of papers to be excluded, number of papers per worker). 
We discuss how quality and cost vary later in the paper when we compare and discuss algorithms.


\subsection{Multi-Run Strategy by Criteria}
\begin{algorithm}[t!]
	\caption{\textbf{M-Runs Algorithm}}
	\hrulefill
	\label{alg:m_runs}
	\begin{tabbing}
	\hspace{\algleftmarginwidth}\={\bf Input:} Items $I$,  Criteria $C$, loss ratio $lr$\\
	\hspace{\algleftmarginwidth}\={\bf Output:} Classified items $CI$\\
     
	\end{tabbing}
    \begin{flushleft}
	\begin{algtab}
    	\hspace{5mm} $CI \leftarrow \{\}$, $UI \leftarrow I$, $thr = \frac{lr}{lr + 1}$, $I^0=$100  randomly selected papers from $I$ \\
        \hspace{5mm} \textit{\textbf{\# Baseline iteration (Run 0)}}\\
        \hspace{5mm} $V^{0} \leftarrow $ collect $J$ votes on $I^0$ for all criteria $C$\\
        \hspace{5mm}  $CI^{0} \leftarrow classify\_items(V^{0}, thr)$\\
        \hspace{5mm} $CI \leftarrow CI \cup CI^{0},\hspace{2mm} UI \leftarrow UI - CI^{0}$\\
        \hspace{5mm} \algforeach{$c \in C$}
        \hspace{5mm} $\hat{\theta}_c \leftarrow \frac{\sum\limits_{i \in I^0} P(i \in OUT_C)}{\left\vert I^0 \right\vert} ,\hspace{2mm} \hat{\alpha}_c = \frac{\sum\limits_{w \in W^0_c} \alpha_{w,c}}{\left\vert W^0_c \right\vert}$   \hspace{1.5mm}  ($W^0_c$ is the set of workers who passed test questions and provided at least one label in baseline iteration)\\
        \algend
             \hspace{5.5mm} \textbf{\#Ranking} $ criteria\_order \leftarrow estimate\_best\_order(\hat{\theta}, \hat{\alpha})$\\   
        \hspace{5mm} \textit{\textbf{\#M-Runs iterations}}\\ 
        \hspace{5.5mm} \algforeach{$c \in criteria\_order$} 
        \hspace{5.5mm} $V^{c} \leftarrow $ collect $J$ votes on $UI$ on $c$\\ 
        \hspace{5.5mm} $CI^{c}_{out} \leftarrow exclude\_items(V^{0}, thr)$\\
        \hspace{5.5mm} $CI \leftarrow CI \cup CI^{c}_{out},\hspace{2mm} UI \leftarrow UI - CI^{c}_{out}$\\
        \algend
        \hspace{5.5mm} $CI_{in} \leftarrow$ tag $UI$ as \textit{"IN items"}\\
        \hspace{5.5mm} $CI \leftarrow CI \cup CI_{in}$\\
        \hspace{5.5mm} \textbf{return} $CI$
	\end{algtab}
    \end{flushleft}
	\hspace*{\algleftmarginwidth}
	\hrulefill
	\hspace*{\algrightmarginwidth}
\end{algorithm}


The multi-run strategy follows the footsteps of the above-mentioned approaches for query optimization in crowd databases that identify the most selective criteria and query based on those first. The difference here is that we also estimate and consider accuracy (we do not want to query the crowd if this brings high disagreement, as it is less cost-effective), and that we work with a specified loss function and a price vs loss trade-offs that are based on the authors' choice.
The algorithms proceeds as follows.



\textbf{Baseline iteration.} We first estimate power and difficulty via a \textit{baseline iteration} (run $k=0$) on a randomly selected subset $I^0$ of the set of candidate papers $I$, as shown in Algorithm~\ref{alg:m_runs} 
(We will get back later in the paper about identifying how large should $I^0$ be).

In step 4 we classify items and estimate accuracy of each worker with a classification algorithm that also provides accuracy estimates such as TruthFinder (TF) \cite{Dong2013}. 
As TF estimates class probabilities, we estimate the power of a criterion $c$ as the expected value of the probability that the criterion applies, as shown in step 7.
We refer instead to the difficulty of a criteria $c$ through the average workers' accuracy on the given criteria, i.e., the average probability that a user, who passed the test questions, gives correct votes on that criteria. 







\textbf{Criteria ranking.} 
Finding the best ordering is trivial if one criterion is more powerful \textit{and} easier than another. Otherwise, different ordering may lead to price/loss points that are on the Pareto frontier and need to be shown to authors for decision. The number of criteria is often low so that considering permutations of all cases where the ordering is not trivial is tractable.
We do so in step 8, by computing for each ordering the expected price and loss for different values of $N_t$ and $J$.
The computation of price and loss can be done as for the previous algorithm. 
Notice that the ordering is very important: 
given an ordering of criteria $OC={c_0, c_1,..c_n}$, the probability of erroneously excluding an item (probability of false exclusion, or PFE) is the probability of erroneously excluding it in the first round (on $c_0$), plus the probability of correctly including it after $c_0$ but erroneously excluding it after $c_1$, and so on. More formally, denoting with $PFE_c$ the probability of erroneous exclusion when processing criteria $c$ and with $PIN_c$ the probability of classifying a paper as IN on criteria $c$:

\begin{equation}
	\begin{aligned}
		PFE = PFE_0+\sum\limits_{m \in 1,2..n} PFE_{m} \prod_{j=0}^{m-1} PIN_{j}
	\end{aligned}
	\label{formula:pfe}
\end{equation}

PFE therefore decreases with PIN, and in practice it decreases sharply if we screen high power criteria first, given that criteria powers over 30\% are quite common.

\textbf{Crowdsourcing iteration.} 
The algorithm  iterates through the criteria, excluding items (classified again based on TF in step 12).


The results of  M-runs (orange) compared with the baseline single run algorithm (blue) are shown in Figure~\ref{f1}b and ~\ref{f1}c, showing loss and precision vs price for different values of $N_t$ and $J$.
The simulation parameters are the same as previously described.
The savings are of approximately 20\%, and are in general higher if the criteria diversity in terms of power and accuracy is higher.  





\input{shortestrun}

%% file: shortestrun.tex
\subsection{Short Adaptive Multi-Run}

The previous algorithms apply the same strategy to all papers left to classify.
The Short Adaptive  Multi-Run  algorithm (SM for short)(Algorithm~\ref{alg:sm_runs}) defines instead an \textit{individual} strategy for each item to be labeled, aimed at identifying the shortest path to decision. The idea is that as we collect votes we understand more about the statistical properties of the overall SLR task (such as criteria power and difficulty) and also of each specific paper, based on the votes obtained for that paper so far.
Therefore, we can estimate which is the criterion to test next for each paper by maximizing the probability of (correctly) classifying it as out in the next run, and we can even decide to give up on a paper (leaving it in) because we realize it is too hard (or too expensive) for the crowd to reach consensus or because the probability that we will classify it as out are low.
In other words, we aim at excluding the papers for which we can do so cheaply and confidently, and leave the rest to the experts (authors).

At an extreme we would like each run to be composed of one vote on one paper for one criterion (hence the name "short run"). Every time we get a vote we learn something new, and we can use this knowledge to optimize the next vote we ask. In practice a run cannot ask for one vote if we use the basic setup of typical crowdsourcing engines (it would not make sense to take time out of a person to explain a task and a criterion and stop after one vote)\footnote{With ad hoc implementations, either stand-alone or on top of commercial engines, and with fast estimation it might be possible to achieve one-vote runs though the key optimization here lies in the personalized strategy: the most important aspect is not so much asking at most one vote in each run, but asking one vote per paper}.

In the following we introduce SM (see Algorithm~\ref{alg:sm_runs}) by first presenting the intuition behind each step and then showing the related math. 

We begin at iteration 0 with an empty set of classified items, both in and out: $CI^0_{in} \cup CI^0_{out} = \emptyset$.
We assume that authors set thresholds for false inclusion and exclusions, that is, values $\overline{P_{out}}$ and $\overline{P_{in}}$ so that we classify a paper $i$ as out if $P(i \in OUT) \geq \overline{P_{out}}$, and analogously for $P(i \in IN)$.
Notice therefore that in SM the authors set the desired precision (as we will see, possibly at the expense of price and recall, but precision is typically non negotiable in SLR as false exclusions are costly).

\textbf{Baseline estimation}. 
We perform a small baseline run as in the previous approach, to estimate power $\hat{\theta}^0_c$ and difficulty (accuracy) $\hat{\alpha}^0_{c}$ for each criterion (Algorithm 2, step 2).
Experiments have shown us that a baseline of 50 items is often sufficient as an initial estimate (as discussed in the following section), considering also that we revise the estimates as we proceed.

\begin{algorithm}[t!]
	\caption{\textbf{SM-Runs Algorithm}}
	\hrulefill
	\label{alg:sm_runs}

	\begin{tabbing}
	\hspace{\algleftmarginwidth}\={\bf Input:} $I$, $C$,$lr,\overline{P_{out}},\overline{P_{in}}$\\
	\hspace{\algleftmarginwidth}\={\bf Output:} $CI=\{CI_{in},CI_{out} \}$ 
	\end{tabbing}
    \begin{flushleft}
	\begin{algtab}
    	\hspace{5mm} $CI \leftarrow \{\}$, $UI \leftarrow I$, 
         $k 
        \leftarrow 0$\\ 
        \hspace{5mm} \textit{\textbf{\# Baseline iteration (Same as Algorithm 1 baseline)}}\\
        \hspace{5mm} $\rightarrow CI, V^{0}, \hat{\theta^0}, \hat{\alpha^0}$\\
                 \hspace{5.5mm}  \textbf{foreach} {$i \in UI$}: $P^0(i\in IN_c/V^{0}_{i,c}) \leftarrow (1-\hat{\theta}^0_c)$\\
        \hspace{5mm} \textit{\textbf{\#SM-Runs iterations}}\\ 
        \hspace{5.5mm} \algwhile{$UI \not= \varnothing$}

        \hspace{5.5mm} $k \leftarrow k + 1$\\
        \hspace{5.5mm} \algforeach{$i \in UI$}
        \hspace{5.5mm} $c(i) \leftarrow \arg\max_{c \in C} \frac{P(V^{k+1,k+n}_{i,c}=OUT)}{N^{min}_{i,c}}$ \\
        \hspace{5.5mm} check\_stop\_condition\_on\_i\\
        \algend
         \hspace{5.5mm} $I^k \leftarrow$ N items with highest p(i)\\
         \hspace{5.5mm} \algforeach{$i \in I^k$}
        \hspace{5.5mm} $v_{i,c}^{k} \leftarrow$ collect a vote for $c$ on $i$\\
        \hspace{5.5mm} $V_{i}^{k} \leftarrow V_{i}^{k-1} \cup v_{i,c}^{k}$\\
        \hspace{5.5mm} $P^k(i\in IN/V^{k}_{i}) \leftarrow \prod_{c \in C} P(i \in IN_c/V_{i,c}^{k})$\\
        \hspace{5.5mm} $P^k(i\in OUT|V^{k}_{i}) \leftarrow 1 - P(i\in IN/V^{k}_{i})$\\
        \hspace{5.5mm} \algif{$P(i\in IN|V^{k}_{i}) > \overline{P_{in}}$}
        \hspace{5.5mm} $CI_{in} \leftarrow CI_{in} \cup \{i \}$\\
        \hspace{5.5mm} $UI \leftarrow UI - \{i \}$\\
        \algend
        \hspace{5.5mm} \algif{$P(i\in OUT|V^{k}_{i}) > \overline{P_{out}}$}
        \hspace{5.5mm} $CI_{out} \leftarrow CI_{out} \cup \{i \}$\\
        \hspace{5.5mm} $UI \leftarrow UI - \{i \}$\\
        \algend
        \algend
        \hspace{5.5mm} update power as per algorithm 1, step 7\\
        \algend
        \hspace{5.5mm} $CI^{diff\_items} \leftarrow$ tag $UI$ as \textit{"IN items"}\\
        \hspace{5.5mm} $CI \leftarrow CI \cup CI^{diff\_items}$\\
        \hspace{5.5mm} \textbf{return} $CI$
	\end{algtab}
    \end{flushleft}
	\hspace*{\algleftmarginwidth}
	\hrulefill
	\hspace*{\algrightmarginwidth}
\end{algorithm}


\textbf{Exclusion probability estimation}. Here we begin the iterations. Before each run of crowdsourcing we try to identify, for each item, and given the votes $V_{i}$ obtained so far for each paper $i$, which criterion is more likely to \textit{efficiently} filter a paper. 
In other words, we identify for each criterion $c$ the minimal number $N^{min}_{i,c}$ of successive out votes we need so that \textit{if} we add $N^{min}_{i,c}$ to $V_{i}$ (resulting in a "imaginary" set of votes $V'_{i}$) \textit{then}
$P(i \in OUT | V'_{i}) > \overline{P_{out}}$, and therefore we exclude the paper and stop working on it.
Intuitively, for each item we want to select criteria that have a low $N^{min}_{i,c}$ (low number of votes and therefore low cost) and a high probability $P(N^{min}_{i,c})$ of getting those out votes.

Notice that every vote on (paper $i$, criterion $c$) we get will move $P(i \in OUT)$ closer or further away from the threshold $\overline{P_{out}}$. This will change our $N^{min}$ and possibly the selected criterion for the next round. 
The probability of getting an out vote for $(i,c)$ also changes, and it does so more strongly when the accuracy for that criterion is higher.

More formally we proceed as follows.
If we denote with $k$ the number of iterations run thus far, and with $V^k_{i,c}$ the votes obtained in the first k runs, then by applying Bayesian rule we have: 

\begin{equation}\label{eq:bayespin}
\begin{split}
P^k(i \in IN_c | V^k_{i,c})= \frac{P^k(V^k_{i,c} | i \in IN_c ) * (1-\hat{\theta}^{k-1}_c)} {P^k(V^k_{i,c})}
\end{split}
\end{equation}

In the formula, after the first run (k=1), the term $\hat{\theta}^{k-1}_c$ is  the proportion of papers to which criteria $c$ applies, as computed after the baseline. $\hat{\theta}_c$ is then updated after each run.

The two $P^k$ factors on the right side of Equation \ref{eq:bayespin} can be determined as follows, where  $J^c_{i,in}$ denotes the number of items $i$ labeled as \textit{in} for criterion $c$:


\begin{equation}
P^k( V^k_{i,c} | i\in IN_c ) =\binom{J^c_i}{J^c_{i,in}} (\overline{\alpha}_c)^{J^c_{i,in}}*(1-\overline{\alpha}_c)^{J^c_{i,out}}  
\end{equation}

and 
\begin{equation}
\begin{split}
P^k(V^k_{i,c}) =P^k(V^k_{i,c}|i \in IN_c ) * (1-\hat{\theta}^{k-1}_c) + \\ 
P^k(V^k_{i,c}|i \in OUT_c ) * \hat{\theta}^{k-1}_c
\end{split}
\end{equation}

Now that we know how to compute $P^k(i\in IN_c | V^k_{i,c})$ and therefore $P(i \in OUT | V^k_{i,c})$ from Equation \ref{formula:p-out}, we can compute how the exclusion probability changes as we add  n=1,2,.. out votes to  $V^k_{i,c}$ obtaining a set we denote as $V^{k \leftarrow n}_{i,c}$ and stop when $n$ is such that $P(i \in OUT |V^{k \leftarrow n}_{i,c}) > \overline{P_{out}}$. 

To assess the probability of getting $N^{min}_{i,c}$ out votes for criteria $c$ on item $i$ we proceed by first computing the probability that the next vote is out, as follows\footnote{To simplify the presentation here we take a single value for criteria accuracy as opposed to a confusion matrix.} (all probabilities are conditional to the votes obtained thus far $V^k_{i,c}$ ):

$P(v^{k+1}_{i,c}=OUT)$ = $\alpha_c * (1-P^k_i (I \in IN_c)) + (1- \alpha_c)P^k_i (I \in IN_c))$.

We then iterate over this formula for getting the probability for the next out votes, remembering that $P^k_i (I \in IN_c)$ will have changed due to the additional out vote.

 
\begin{figure*}[htb]
	\centering
    \begin{minipage}{1\linewidth}
		\includegraphics[width=1\textwidth]{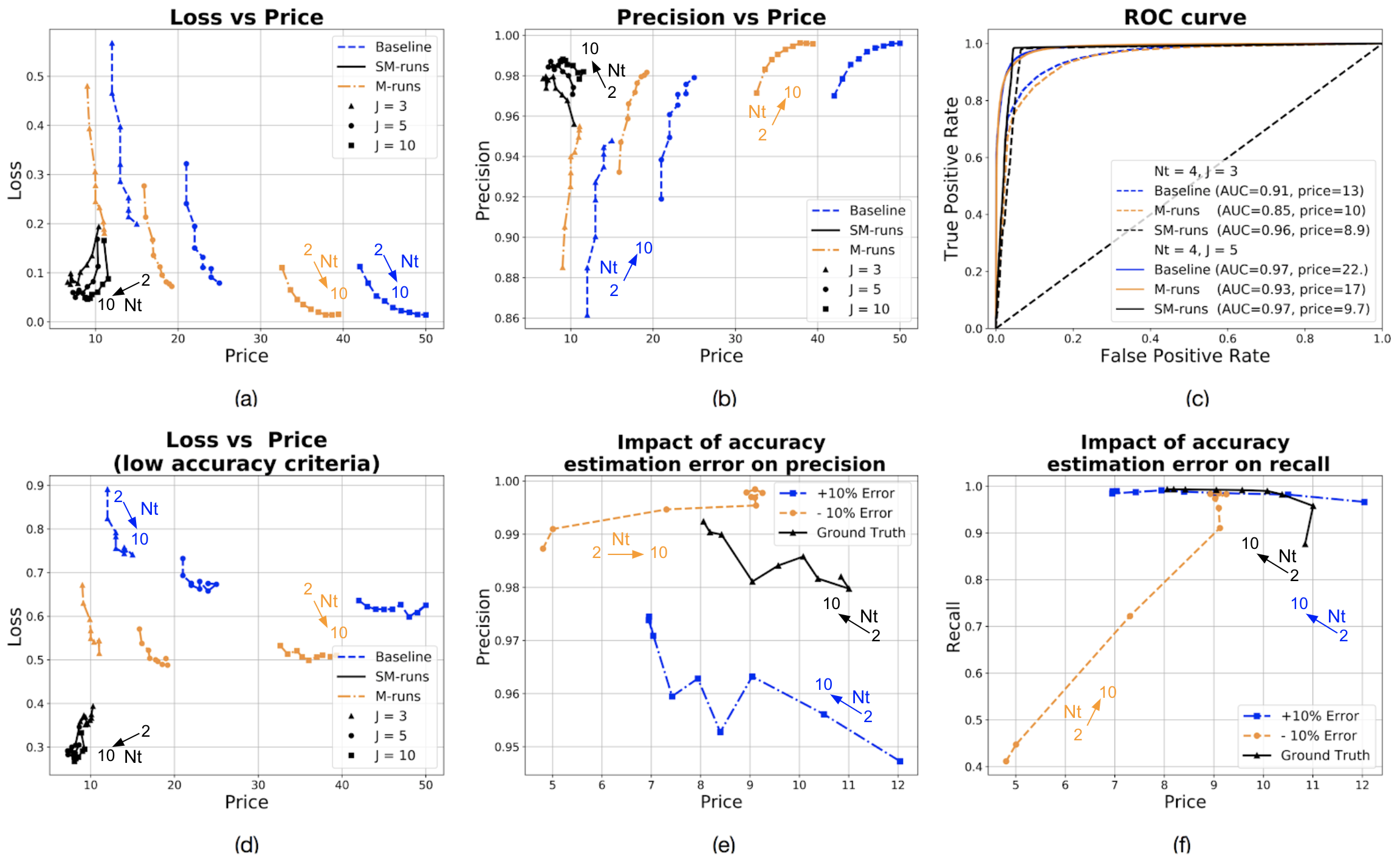}
		\caption{Behavior of algorithms. Charts are simulated with 1000 papers, four criteria of power $=[c1=0.14, c2=0.14, c3=0.28, c4=0.42]$, $Nt=[2,3,..,10]$, $lr=5$. 
 Workers are assumed to be cheaters with probability 0.3, and the rest has uniform accuracy in (0.5-1). Accuracy on OUT papers are 10\% higher, as seen in experiments. See text for description.}   
\label{f1}
	\end{minipage}
\end{figure*}

\textbf{Ranking.} We rank criteria for each item by weighing cost ($N^{min}_{i,c}$) and probability of success (probability $P(V^{k+1,k+n}_{i,c}=OUT)$ of getting $N^{min}_{i,c}$ consecutive out votes). 
We define the value of applying a criterion as the price we have to pay for unit of probability of classifying the item as out in the next $N^{min}_{i,c}$ votes, that is:  $Value_{i,c} =  P(V^{k+1,k+n}_{i,c}=OUT) / N^{min}_{i,c} $
We then borrow ideas from  predicate ranking optimization in query processing \cite{Hellerstein93predicates} that essentially ranks based on selectivity/cost (although here we do so per item and assess it at each iteration). Applying the same logic we look for each paper for the criterion with maximum value: $Value_{i} = \max\limits_{c \in C}  P(V^{k+1,k+n}_{i,c}=OUT) / N^{min}_{i,c} $

In developing SM we explored alternative approaches: a main one we explored involves estimating how $P(i \in OUT)$ is likely to change if we ask for one vote on $c$, as an attempt to drive our choice for which vote to ask next. 
With relatively simple math, we can estimate the probability of the next vote being in or out, and the impact that this has on $P(i \in OUT)$, and we can select the criteria that leads us closer to the threshold. This initial choice however has an undesired behavior: if there is a low accuracy, high power criterion, it leads us to choosing this criterion. 
However, the low accuracy means we only take little steps towards our threshold, making the walk long and expensive. Instead, we choose criteria that can provide large variations towards the out threshold.

\textbf{Stopping.} 
As we iterate, we can see that $Value_{i}$ may be so low (for example, if we get conflicting votes) that it becomes ineffective to poll the crowd for that item.
We can therefore stop working on papers for which $Value_{i}$ is lower than a threshold based on authors' preferences (Notice that we disregard the money already spent on a paper, as that is a \textit{sunk cost} \citep{Arkes85sunkcost}).
The reasonable threshold here depends on the cost ratio $cr$ of the crowd cost for a single vote on one paper and criterion (PPL from Formula \ref{ppp}) divided by the author classification cost. The lower the cost ratio, the more convenient it is to insist with the crowd. 
For typical cost ratios, considering classification costs as estimated in the literature (see, e.g., \cite{Mortensen2016crowd}) of around 2\$ per abstract (for the US, in the medical field and including overhead), a good empirically set threshold is 100. we do not discuss this threshold further here but refer the interested reader to \textit{http://jointreserch.net} for details.

\textbf{Crowdsourcing iteration.} 
Ranking determines the priority for the next batch of votes. 
The batch size is the minimal size that can practically be achieved while ensuring each worker gets value for the time they spend learning and doing the task. 
In practice, it rarely makes sense to offer batches of less than 10 items as they are less attractive. 
We return to the crowd to ask one more vote for each paper in the batch, determine the probability of exclusion as discussed above and classify paper as out if $P(i \in OUT) > \overline{P_{out}}$.
If there are no more paper left to classify we stop, else we iterate.

We next analyze the results of the algorithm and discuss its properties, also in light of crowdsourcing experiments.

%% file: experiments.tex
\section{Analysis and Experiments}


\textbf{Simulations.} 
We first show the behaviors of  algorithms via simulations.
The strategies presented here have a number of parameters and the behavior varies in interesting ways as we change these parameters. Here we limit to point out some aspects we found particularly interesting and provide an in-depth analysis online for the interested reader, along with the code to replicate both simulation and analysis of experiments\footnote{http://jointresearch.net}. 
We also remind that authors do not set or estimate any parameter: they simply need to state their loss function and the preference for given loss vs price points when there is no Pareto-optimal value.

Figure \ref{f1} shows the result of a simulation run with 1000 papers and parameters as described in the caption.
It plots the loss vs price curve for the SM strategy for the same  scenario discussed for the other algorithms (The SM variant adopted here has a 1000 papers run, assumes a stopping threshold of 100, and shows an average of 50 simulations). $\overline{P_{out}}$ is 0.99.
Figure \ref{f1}(a) and (b) show that SM can achieve the same loss and precision for a fraction of the cost (both could improve by changing $\overline{P_{out}}$, though increasing the price)\footnote{We omit plotting the std bars as they would make the chart unreadable}.
Notice that price and loss both decrease (at least initially) as we increase the number of tests $N_t$, which is our "knob" to increase accuracy (and cost) of workers.
This is because SM detects the increased accuracy and adapts to it by asking for less votes for the same loss and precision.
Figure~\ref{f1}c shows the ROC curve where we can see that SM has a greater area for a much smaller price.
Charts are analogous in terms of shapes and trends for other values of $\theta,J$ (and for $N_t$ as well for the ROC curve) so we do not show them.
Figure~\ref{f1}d shows again loss vs price but this time assuming the presence of a very difficult  criterion (accuracy of 0.55), and shows the robustness of SM on the loss (even with the very conservative loss ratio of 5 we used here).

\begin{figure}[hbt]
	\centering
    \begin{minipage}{1\linewidth}		\includegraphics[width=0.9\textwidth]{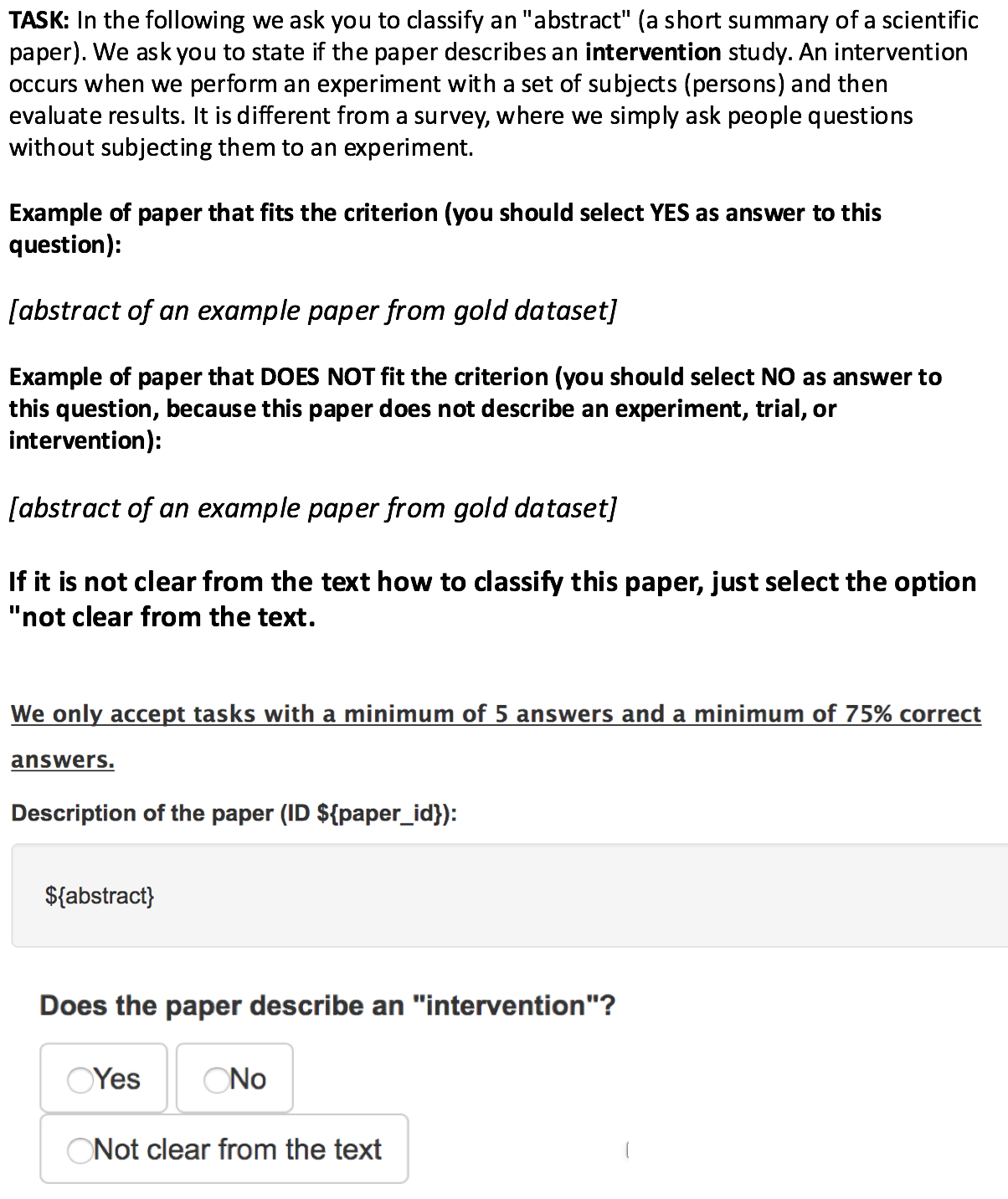}
		\caption{Classification task for SLR}   
\label{task}
	\end{minipage}
\end{figure}



Figure~\ref{f1}e and ~\ref{f1}f show the impact of estimation error for accuracy on precision and recall respectively. Notice that if we underestimate (orange line) we achieve higher precision (we are more conservative). For recall, if we underestimate accuracy and accuracy is low ($N_t$ is low) then we get very low recall: we give up rather early, leaving papers to authors to classify. As accuracy increases, the differences smooth out and are within the variance. 
The charts for power estimation error have essentially  the same shape and are not shown.

Baseline runs and numbers of labels per paper affect the estimation errors. The issue is not so much the number of papers in the baseline: 40-50 papers suffice to estimate power within a 5-7\% margin of error (consider the problem similar to estimating the fairness of a coin modeled as a Beta distribution, 50 tosses would give a reasonable estimate).
Furthermore, estimates are re-assessed as we go.
The key here is rather to enable a good accuracy estimation, and experiments have shown that with less than 3 votes per paper the estimation error grows above 10\% and with low accuracy criteria this can generate very low recall (as we may believe accuracy to be at 0.5).
Experiments with variations of the stopping threshold also produced limited effect. When going from 100 to 150, the recall increased by approximately 0.04\%, always keeping the precision threshold at 0.99. The price difference is also negligible.

\begin{figure*}[hbt]
	\centering
    \begin{minipage}{1\linewidth}
		\includegraphics[width=1\textwidth]{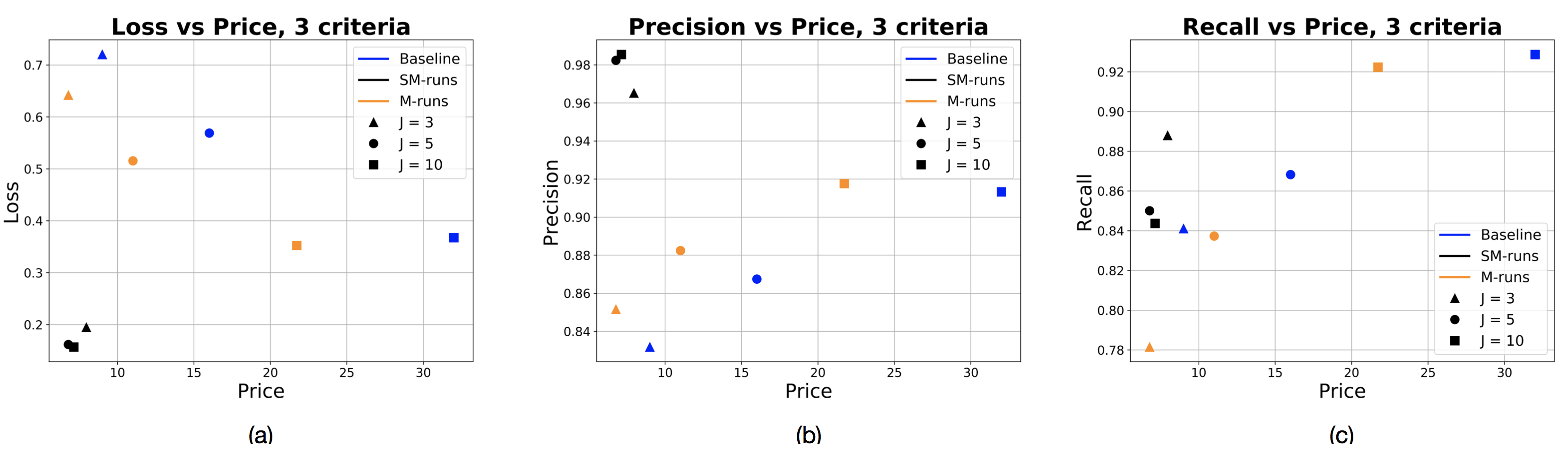}
		\caption{Behavior of SM with data from experiments. Runs of 1000 papers, accuracy and power as described in the text, the other parameters are unchanged.}   
\label{f2}
	\end{minipage}
\end{figure*}

\textbf{Experiments.} Between January and September 2017 we performed a set of studies and experiments on two commercial crowdsourcing engines (CrowdFlower and Mechanical Turk). 
We ran a total of 20 experiments with different settings, asking workers to label a total of 174 papers with two to four exclusion criteria (a total of 514 classification decisions) taken from two systematic reviews, one done by us in an interdisciplinary area (computer science and social sciences) reviewing technology for fighting loneliness (reference omitted for double blind), and the other in medicine \cite{veronese2017weight} having more complex exclusion criteria. 
We collected votes by over 3200 respondents. 
These initial studies helped us to understand the nature of the problem, estimate crowd accuracies, get a feeling for latencies and costs, and also refine task design which, although orthogonal to our goals here, is important for getting good results \cite{yang2016modeling}.
In the following we focus on the experiments to assess the validity of SM with respect to other algorithms and baselines.



\textit{Setup.} To this end we classified 374 papers on AMT by posting tasks that were asking crowd workers to classify many papers based on one criteria. We requested workers with an HIT approval rate of 70\% or more. 
The task starts by explaining the criteria to workers, providing a positive and a negative example, and then asking to label the papers as in, out, or unclear from the text (Figure~\ref{task}). Adding the latter option was a result of previous experiments where many workers complained that this option was missing and were unsure about what to answer to qualify for payment.
We also informed workers that they would need to get 75\% of the answers as correct in order to be paid.
The examples are taken from the "gold" dataset, which are the papers classified by researchers in our team. 
Each worker saw the same example papers in the instructions for the same criteria. 
The three criteria we tested involved assessing whether the paper included an \textit{intervention} (as shown in Figure~\ref{task}), whether it described studies on adults \textit{65 and older}, and whether it involved the use of \textit{technology}.

The task proceeded by criteria, not by paper: we showed instruction for a criterion, and then asked for classification on that criterion. We chose this option as we noticed in initial experiments that explaining and understanding criteria takes effort, and teaching workers the subtleties of several criteria at the same time may lead to increased effort and reduced accuracy.  
Workers could classify the papers until they wished to do so. 
We repeated the  process for the three criteria getting at least 5 votes per paper (we collected up to 15 votes per paper for the intervention criterion show in the figure as it had low accuracy and we wanted to analyze it more deeply). 
Before running the task we did some pre-runs to assess the proper pay. In terms of costs, we experimented different payments, always making sure that we stay well over 8USD per hour based on estimated completion times, which results in approximately 10cents per vote. 
We did not screen workers with test questions, though the experiments gave us a dataset over which we can now use to "simulate" the effect of filtering out workers that did not get 100\% accuracy on the first $N_t$ questions.
The dataset is publicly available online at \textit{jointresearch.net}.


\textit{Results.} 
All tasks were completed in a few hours, and we assume that if we had more papers they would have been classified with sublinear increment in time. 
As expected (see table below), and consistently with the literature ~\cite{Mortensen2016crowd},  power and workers' accuracy vary significantly by criteria. Use of technology has high power as words related to technology are very common and it is hard for keyword-based queries to filter for the specific use of technology we look for.

\bigskip

\begin{tabular}{ l c r r}
\hline
 Criteria: &  \textit{intervention} & \textit{use of tech} & \textit{65 and older} \\
 Power & 0.24 & 0.61 & 0.05 \\
  Accuracy & 0.60 & 0.77 & 0.75 \\
  \hline
\end{tabular}

\bigskip


The presence of a criteria (intervention) with rather low accuracy underscores the importance of an adaptive approach where we focus on high-accuracy criteria and leave the leftover papers to the authors.

Another interesting finding was that nearly half of the papers erroneously classified by the crowd were either errors in the gold data (our error) or they were cases where after reading in detail the abstract we were unsure ourselves. This prompted us to study a bit deeper average agreement among expert raters. 
Mateen et al report on an experiment that measured agreement on around 96\% of the papers \cite{Mateen_titlevsabstract_2013}. 
An analysis of SLRs conducted by our team reported 92\% agreement among two raters and, in addition, 2\% of cases where one rater was unsure.
This indicates that the $\overline{P_{out}}$ precision threshold we picked of 0.99 is in line or even exceeding current standards.
We also observed that the mere act of having the need to explain criteria to others (that also demand fairness in job acceptance) forced us to be very precise and indeed, looking back at our own classification, we found errors also due to a certain imprecision in the initial definitions.


Using the experimental data to fuel simulations did not bring significant changes to the charts already discussed, although there are interesting differences and we focus on these in the following, especially to underline the limits of SM.
One interesting aspect is that actual data do not precisely and consistently fit the model: workers accuracy cannot always be  modeled as i.i.d. variables, and the margin of error in predicting future accuracy from past is rather high, even if we vary the testing patterns. 
This is not entirely surprising as some workers may improve as they proceed with the task while others may get sloppy or tired, and indeed optimal testing to cater for these issues is an active area of research ~\cite{Bragg2016}.

Figure~\ref{f2} shows the results of such experiments-fueled simulations, assuming no additional tests filtering. 
Figure~\ref{f2}a shows the usual loss vs price chart, and the results are fairly consistent with the simulations.
Figures~\ref{f2}b and c break this down by precision and recall. 
The latter is particularly interesting as the recall for baseline is somewhat comparable to M-runs, or at least it makes for a non obvious choice for lower level of $J$. This is probably due to the relatively low accuracy we have in these (relatively untested) set of workers.

Finally we observe that in terms of overall cost, even at 10cents per vote we remain well below author cost for the paper we screen out (from 20 to 40\%).



%% file: conclusion.tex
\section{Conclusion}
The SM algorithm seems to have the potential to outperform baselines for finite pool classification problems, and especially for SLR. We also confirm initial findings that crowdsourcing is feasible for paper screening in SLR. We have also explored extensions of this approach to general classification problems, including problems combining crowd and machine classification~\cite{Krivosheev_www18poster}.
The work still has many limitations, especially that of improving the estimation of accuracies and extending the model to cover the case where workers' accuracy is very "noisy". 

\textbf{Acknowledgements.} This project has received funding from the EU Horizon 2020 research and innovation programme under the Marie Skodowska-Curie
grant agreement No 690962. 
